\documentclass[conference]{IEEEtran}
\IEEEoverridecommandlockouts
\usepackage{cite}
\usepackage{amsmath,amssymb,amsfonts}
\usepackage{algorithmic}
\usepackage{graphicx}
\usepackage{textcomp}
\usepackage{xcolor}
\usepackage[pdftex,
pdfauthor={Adel Rahimi},
pdftitle={Improving Information Retrieval  Results for Persian Documents using FarsNet},
pdfsubject={Improving Information Retrieval  Results for Persian Documents using FarsNet},
pdfproducer={adel},
pdfcreator={Mac}]{hyperref}
\graphicspath{ {images/} }
\def\BibTeX{{\rm B\kern-.05em{\sc i\kern-.025em b}\kern-.08em
    T\kern-.1667em\lower.7ex\hbox{E}\kern-.125emX}}
\begin{document}

\title{Improving Information Retrieval  Results for Persian Documents using FarsNet\\}

\author{\IEEEauthorblockN{1\textsuperscript{st} Adel Rahimi}
\IEEEauthorblockA{\textit{Languages and Linguistics Center} \\
\textit{Sharif University of Technology}\\
Tehran, Iran \\
adel.rahimi72@student.sharif.edu}
\and
\IEEEauthorblockN{2\textsuperscript{nd} Mohammad Bahrani}
\IEEEauthorblockA{\textit{Languages and Linguistics Center} \\
	\textit{Sharif University of Technology}\\
	Tehran, Iran \\
	bahrani@sharif.edu}
}

\maketitle

\begin{abstract}
In this paper we propose a new method for query expansion, which uses FarsNet (Persian WordNet) to find similar tokens related to the query and expand the semantic meaning of the query. For this purpose, we use synonymy relations in FarsNet and extract the related synonyms to query words. This algorithm is used to enhance information retrieval systems and improve search results. The overall evaluation of this system in comparison to baseline method (without using query expansion) shows an improvement of about 9 percent in Mean Average Precision (MAP).
\end{abstract}


\section{Introduction}
In Information Retrieval (IR), the searched query has always been an integral part. When a user enters a query in the information retrieval system the keywords they use might be different from the ones used in the documents or they might be expressing it in a different form. Considering this situation, the information retrieval systems should be intelligent and provide the requested information to the user. According to Spink (2001), each user in the web uses 2.4 words in their query; having said that, the probability of the input query being close to those of the documents is extremely low [22]. The latest algorithms implement query indexing techniques and covers only the user's history of search. This simply brings the problem of keywords mismatch; the queries entered by user don't match with the ones in the documents, this problem is called the lexical problem. The lexical problem originates from synonymy. Synonymy is the state that two or more words have the same meaning. Thus, expanding the query by enriching each word with their synonyms will enhance the IR results.

This paper is organized as follows. In section II, we discuss some previous researches conducted on IR. In section III, the proposed method is described. Section IV, represents the evaluation and results of proposed method; and finally, in section V, we conclude the remarks and discuss some possible future works.

\section{Previous Works}
One of the first researchers who used the method for indexing was Maron (1960) [11]. Aforementioned paper described a meticulous and novel method to retrieve information from the books in the library. This paper is also one of the pioneers of the relevance and using probabilistic indexing. Relevance feedback is the process to involve user in the retrieved documents. It was mentioned in Rocchio (1971) [15], Ide (1971) [8], and Salton (1971) [19]. In the Relevance feedback the user's opinion for the retrieved documents is asked, then by the help of the user's feedbacks the relevance and irrelevance of the documents is decided. In the later researches, relevance feedback has been used in combination with other methods. For instance, Rahimi (2014) [14] used relevance feedback and Latent Semantic Analysis (LSA) to increase user's satisfaction. Other researches regarding the usage of relevance feedback are Salton (1997) [18], Rui (1997) [16], and Rui (1998) [17].

In the next approaches, the usage of thesauri was increased. Zazo used thesauri to "reformulate" user's input query [23]. Then came the WordNet. WordNet was one the paradigm shifting resources. It was first created at Princeton University's Cognitive Science Laboratory in 1995 [12]. It is a lexical database of English which includes: Nouns, Adjectives, Verbs, and Adverbs. The structure of WordNet is a semantic network which has several relations such as: synonymy, hypernymy, hyponymy, meronymy, holonymy, and etc. WordNet contains more than 155,000 entries. Using WordNet for query expansion was first introduced in Gong (2005) [5]. They implemented query expansion via WordNet to improve one token search in images and improved precision. Another research conducted by Pal (2014) showed that the results from query expansion using standard TREC collections improves the results on overall [13]. Zhang (2009) reported 7 percent improvement in precision in comparison to the queries without being expanded [24]. Using WordNet for query expansion improved 23 to 31 percent improvement on TREC 9, 10, and 12 [10].

Liu (2004) used a knowledge database called ConceptNet which contained 1.6 million commonsense knowledge [9]. ConceptNet is used for Topic Gisting, Analogy-Making, and other context-oriented inferences. Later, Hsu (2006) used WordNet and ConceptNet to expand queries and the results were better than not using query expansion method [6].

FarsNet [20] [21] is the first WordNet for Persian, developed by the NLP Lab at Shahid Beheshti University and it follows the same structure as the original WordNet. The first version of FarsNet contained more than 10,000 synsets while version 2.0 and 2.5 contained ~20,000 synsets. Currently, FarsNet version 3 is under release and contains more than 40,000 synsets [7].

\section{Proposed Method}
Each word in FarsNet has a Word ID (WID). Each WID is then related to other WIDs e.g. words and their synonyms are related to each other in groups called synsets.

As mentioned before, often the user input doesn't match with the ones used in the documents and therefore the information retrieval system fails to fulfil user's request. Having said that; the present paper utilizes FarsNet and its synonymy relations to use in query expansion.

We use the original synsets of FarsNet 2.5 as dataset. However, the data is first cleaned and normalized. Normalization refers to the process where the /ی/ is replaced with Unicode code point of 06CC and /ک/ is replaced by Unicode code point of 06A9. 

The input of the algorithm is the string of input queries. Then the input string is tokenized. Tokenization is the process of separating each word token by white space characters. In the next step, each token is searched in FarsNet and if it is found, the WID of the token will be searched in the database of synonyms; in other words, FarsNet Synsets. Finally, each word is concatenated to its synonyms and they are searched in the collection. Snippet below shows the pseudo code of the query expansion method.

Sample input and output are:
\newline

Input: [Casualties \ of drought]

\begin{center}
\includegraphics[scale=0.5]{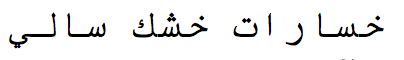}

\end{center}

Output: [drought Casualties \ waterless \ dry \ dried up]

\begin{center}

\includegraphics[scale=0.5]{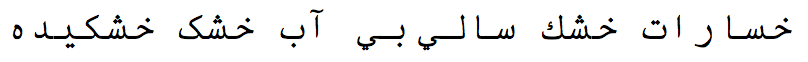}
\end{center}
بي آب خشک خشکيده خسارات خشك سالي

\begin{verbatim}
GET input_query
L <- an empty list

FOR token IN input_query:
Wid  <- find token's WID in FarsNet
INSERT(Wid , L)

Expanded_Query <- inputـquery
FOR wid IN L:
Syns <- find synonym of wid in Synset
CONCAT(Expanded_Query, Syns)

Search Expanded_Query in Collection
END

\end{verbatim}

\section{Experimental Results}
In the evaluation phase, we used Hamshahri Corpus [2] which is one of the biggest collections of documents for Persian, suitable for Information Retrieval tasks. This corpus was first created by Database Research Group at Tehran University. The name Hamshahri comes from the Persian newspaper Hamshahri, one of the biggest Persian language newspapers. Hamshahri corpus contains 166,000 documents from Hamshahri newspaper in 65 categories. On average, each document contains 380 words and in general the corpus contains 400,000 distinct words. This corpus is built with TREC standards and contains list of standard judged queries. These queries are judged to be relevant or irrelevant to the queries based on real judgments. The judgment list contains 65 standard queries along with the judgements and some descriptions of the queries. Sample queries include:
\begin{center}
[women basketball]

[teaching gardening flower]

[news about jungles' fires]

[status of Iran's carpet export]

[air bicycle]
\end{center}
\begin{center}

\includegraphics[scale=0.4]{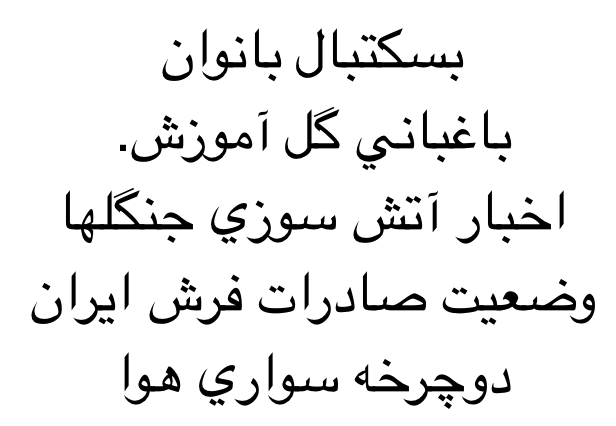}
\end{center}

In the present paper, the information retrieval experiments are based on standard queries of Hamshahri corpus.

For assessment of the proposed algorithm in a real information retrieval situation we used Elasticsearch database [1]. Elasticsearch is a noSQL database which its base is document, hence called document based database. Elasticsearch uses Lucene as its engine. The evaluation process started with normalizing all the documents in Hamshahri corpus. Then some articles that were incomplete or had errors were removed so that they could be indexed in Elasticsearch. In the end, the total number of 165,000 documents were indexed in Elasticsearch. Code snippet below shows a sample of index structure in Elasticsearch database.

\begin{verbatim}
_index: "Hamshahri" [Default-Elasticsearch Index]

_type: "articles" [Default-All our types are Hamshahri document]

_id : "AV9Np3YfvUqJXrCluoHe" [random generated ID]

DID: "1S1" [Document ID in Hamshahri Corpus]

Date: "75\\04\\02" [Document date in Iranian Calendar, \\ is 
for character escape]

Cat: "adabh" [Document category e.g. adab-honar]

Body: "…" [Document body]

\end{verbatim}

We arranged two sets of experiments for evaluation of the algorithm: without query expansion (baseline) and with query expansion (proposed). First, for each query in the standard query list of Hamshahri corpus, we searched in Elasticsearch database and retrieved the results. In the next step, we expanded each query using proposed method and searched each expanded query in Elasticsearch.

In order to evaluate the precision of the retrieved documents in each experiment, we used "TREC\_Eval" tool [3]. TREC\_Eval is a standard tool for evaluation of IR tasks and its name is a short form of Text REtrieval Conference (TREC) Evaluation tool. The Mean Average Precision (MAP) reported by TREC\_Eval was 27.99\% without query expansion and 37.10\% with query expansion which shows more than 9 percent improvement.

\begin{figure*}[tbp]
	\centering
	\includegraphics[width=0.70\linewidth]{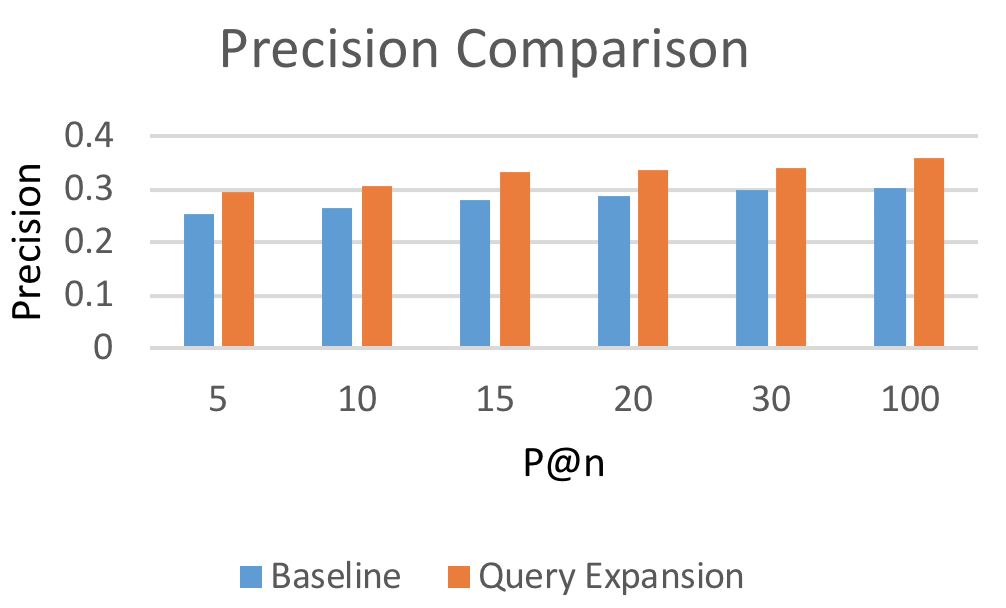}
	\caption{Precision comparison for different numbers of retrieved documents in two sets of experiments}
\end{figure*}

Table 1 and Figure 1 show the precision at the first n retrieved documents (P@n) for different numbers of n in two sets of experiments. In all P@n states the precision of Query Expansion algorithm was higher than the baseline.


\begin{figure*}
	\centering
	\includegraphics[width=0.7\linewidth]{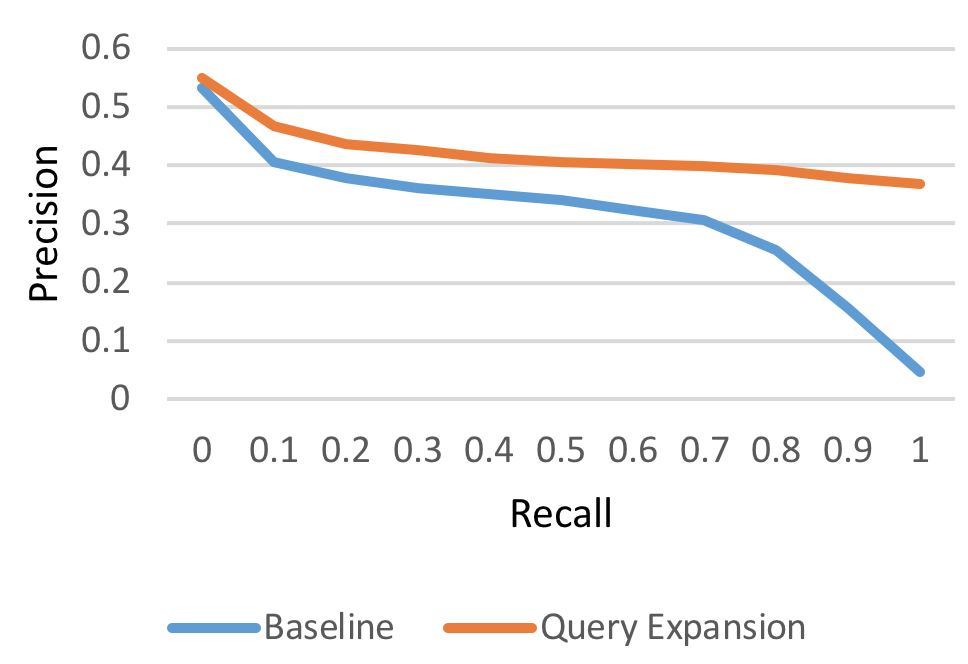}
	\caption[Plot of precision vs recall for two sets of experiments]{Plot of precision vs recall for two sets of experiments}
	\label{fig:precisionvsrecall2}
\end{figure*}

Figure 1 shows the plot of precision vs recall for two sets of experiments. This plot shows that our method will improve the overall quality of Information Retrieval system.
\begin{table}
\begin{center}
	
	\caption{Precision at the first n retrieved documents (P@n) for different numbers of n. ``Plus" denotes with query expansion and ``Minus" denotes without query expansion}
	\begin{tabular}{p{2cm}p{2cm}p{2cm}}
		\hline\noalign{\smallskip}
		P   &   Minus & Plus  \\
		\noalign{\smallskip}
		\hline
		\noalign{\smallskip}
		P@5 & 0.2523 &  0.2954\\
		P@10 & 0.2646 & 0.3076  \\
		P@15 & 0.2800 & 0.3322 \\
		P@20 & 0.2862 & 0.3377 \\
		P@30 & 0.2989 & 0.3415 \\
		P@100 & 0.3044 & 0.3610 \\
		\hline
	\end{tabular}
\end{center}
\end{table}

\section{Conclusions}

In this paper, we proposed a method for query expansion in IR systems using FarsNet. Results from this approach showed about 9\% improvement in Mean Average Precision (MAP) for document retrieval. 

In the future researches, we will use FarsNet 3.0 and also, we will modify and revise some synsets in the FarsNet, in order toincrease the precision for Information Retrieval.

\end{document}